\documentclass[pra,showpacs,twocolumn]{revtex4}
\usepackage{bm,graphicx,amsmath}
\usepackage{amsfonts}
\usepackage{amssymb}
\usepackage{amsthm}
\begin{document}
\title{Cavity QED nondemolition measurement scheme using quantized atomic motion}
\author{Jonas Larson$^1$}
\email{jolarson@kth.se}
\author{Mahmoud Abdel-Aty$^2$}
\affiliation{$^1$NORDITA, 106 91 Stockholm, Sweden\\
$^2$ College of Science, University of Bahrain, 32038 Kingdom of Bahrain}
\date{\today}
\begin{abstract}
Considering ultracold atoms traversing a high-$Q$ Fabry-Perot cavity, we theoretically demonstrate a quantum nondemolition measurement of the photon number. This fully quantum mechanical approach may be understood utilizing concepts as effective mass and group velocity of the atom. The various photon numbers induce a splitting of the atomic wave packet, and a time-of-flight measurement of the atom thereby reveals the photon number. While repeated atomic measurements increase the efficiency of the protocol, it is shown that by considering long interaction times only a few atoms are needed to resolve the photon number with almost perfect accuracy.  
\end{abstract}
\maketitle

\section{Introduction}
Any measurement of a quantity $\hat{A}$ of a quantum system has an impact on the system itself. For example, exactly determining $\hat{A}$ implies that any knowledge of a conjugate variable is lost. Moreover, the detection might even be destructive by nature, {\it e.g.} the standard way of measuring the electromagnetic field is by photocounting detectors where photons are actually absorbed. However, there exist situations where a measurement of some quantity $\hat{A}$ does not induce any back-action quantum noise on $\hat{A}$. That is, provided $\hat{A}$ is a constant of motion, then a second measurement of $\hat{A}$, after some time-delay relative the first measurement, would reveal the same value as obtained in the first measurement. This is called a quantum nondemolition (QND) measurement~\cite{qnd}. 

By now, QND measurements of the optical field have been demonstrated both for an optical fiber~\cite{qndexp1} and in cavity QED setups~\cite{qndexp2}. In the cavity experiments to date, an atomic Ramsey interferometer technique has been utilized. The different phases aquired for an atom interacting dispersively with the cavity field either in its excited state $|e\rangle$ or ground state $|g\rangle$ contain information about the photon number $n$. Repeating the atomic QND measurements sufficiently many times determines the photon number definitely.  

The kinetic energy of the atoms in these experiment greatly exceeds the atom-field interaction energy, and thereby they can be safely treated by classical means. For ultracold atoms on the other hand, mechanical effects induced by the light fields become important. Such actions are indeed the building blocks for cooling and trapping of neutral atoms~\cite{atcol}. In the cavity QED community, it has long been known that the atom-field dynamics may be considerably modified by treating the atomic motion quantum mechanically together with taking spatial mode variations into account~\cite{mazer}. The system now includes additional degrees of freedom that become correlated with the cavity field. Already back in 1989 it was shown that deflection of a beam of ultracold atoms resonantly interacting and transversely scattered from a quantized standing wave field will depend on the actual photon distribution of the field~\cite{stig}. For a classical field, the same effect, named optical Stern-Gerlach, was presented in Ref.~\cite{optSG}. These observations led to the idea of performing QND measurments of the cavity field using ultracold atoms dispersively interacting with the field. Reference~\cite{qndquant1} considered a beam of ultracold atoms transversely passing a Fabry-Perot cavity. In the Raman-Nath regime, valid for very short interaction times, it was shown that the deflection of the atomic beam after interacting with the cavity field depends on the number of photons. Thus, by recording the atomic positions of a sequence of atoms having interacted with the cavity field, a QND measurement of the photon number is possible.

In this paper we address a different QND measurement where the time-of-flight of the atom is recorded. Instead of studying the deflection of transversely passing atoms, we consider atoms traversing the Fabry-Perot cavity along its axial direction. Our treatment is fully quantum mechanical, taking into account for; the atomic scattering effects occurring as the atom enters and exits the cavity, the quantum pressure arising from the atomic kinetic energy term (hence going beyond the Raman-Nath regime), as well as the uncertainty of the velocity of the incoming atom. The scheme is first studied in a semi-analytical model which relies on the concepts of effect mass and group velocity for the atom. By imposing a sort of single-band approximation, the semi-analytical model demonstrates that a single atom is sufficient for performing the QND measurement in the ideal situation provided the effective interaction time is long enough. The full system, going beyond single-band and taking into account for a finite cavity, is considered numerically, and it is found that typically around ten atoms is enough for achieving a highly efficient QND measurement.

\section{Model system}\label{sec2}
We consider an ultracold two-level atom sent through a Fabry-Perot
cavity along its axial axis. The internal ground and excited
atomic states are labeled $|g\rangle$ and $|e\rangle$, and their energy
difference $\hbar\Omega$. The atom interacts with a single cavity
mode with frequency $\omega$. Moving to an interaction frame and
imposing the rotating wave approximation, the Hamiltonian reads
\begin{equation}
\hat{H}'=\frac{\hat{P}^2}{2m}+\frac{\hbar\tilde{\Delta}}{2}\hat{\sigma_z}+\hbar \left[g(\hat{X})\hat{a}^\dagger\hat{\sigma}^-+g^*(\hat{X})\hat{\sigma}^+\hat{a}\right].
\end{equation}
Here, $\hat{X}$ and $\hat{P}$ are the atomic center-of-mass position
and momentum respectively, $\tilde{\Delta}=\Omega-\omega$ is the
atom-field detuning, $g(\hat{X})$ the effective position dependent
coupling, $\hat{a}^\dagger$ ($\hat{a}$) the creation (annihilation)
operators for the field, and the Pauli-operators are
$\hat{\sigma}_z=|e\rangle\langle e|-|g\rangle\langle g|$,
$\hat{\sigma}^+=|e\rangle\langle g|$, and
$\hat{\sigma}^-=|g\rangle\langle e|$. For a
cavity of length $L$ we have
\begin{equation}\label{pot1}
g(\hat{X})=\left\{\begin{array}{lll}\tilde{g}_0\cos(k\hat{X}) & 0\leq \hat{X}\leq L\\
0 & \mathrm{othervice}.\end{array}\right.
\end{equation}
The length $L$ of the cavity is assumed to be much larger than the wavelength of the field, and hence, the Hamiltonian is quasi-periodic. $k$ is the wave number
and $\tilde{g}_0$ the effective atom-field coupling. Letting  $E_{2r}=\frac{\hbar^2k^2}{m}$ define the
characteristic energy, we scale the parameters accordingly
\begin{equation}
\begin{array}{ll}
g_0=\tilde{g}_0/E_{2r}, & 
\Delta=\tilde{\Delta}/E_{2r}, \\ \\ t=\tilde{t}E_{2r}/\hbar, &
\hat{x}=k\hat{X},
\end{array}
\end{equation}
where $\tilde{t}$ is the unscaled time. Using the fact that within
the rotating wave approximation, the number of excitations
$\hat{N}=\hat{a}^\dagger\hat{a}+\hat{\sigma}_z/2$ is preserved, 
the Hamiltonian may be written on block form within the states
$|n,g\rangle$ and $|n-1,e\rangle$, $|n\rangle$ being the Fock
state with $n$ photons. Thus, we have
$\hat{H}'=\hat{H}_0'\otimes\hat{H}_1'\otimes\hat{H}_2'\otimes...$ where
$\hat{H}_0'=-d^2/2d\hat{x}^2-\Delta/2$ and 
\begin{equation}\label{ham1}
\hat{H}_n'=-\frac{1}{2}\frac{d^2}{d\hat{x}^2}+\left[\begin{array}{cc}
\frac{\Delta}{2} & g_0\cos(\hat{x})\sqrt{n} \\ \\
g_0\cos(\hat{x})\sqrt{n} & -\frac{\Delta}{2}\end{array}\right].
\end{equation}
Here we have taken $g_0$ to be real.

For ultracold atoms in the dispersive regime, $\Delta\gg g_0\sqrt{n}$, 
we adiabatically eliminate the excited atomic level $|e\rangle$ to
obtain a Hamiltonian describing the dynamics for the atomic ground state
alone. Following standard procedures \cite{adel}, one derives
\begin{equation}\label{ham2}
\hat{H}_n=-\frac{1}{2}\frac{d^2}{d\hat{x}^2}+Un\cos^2(\hat{x}),
\end{equation}
where $U=\frac{g_0^2}{\Delta}$ is the amplitude of the single photon dipole
induced potential.

\section{Semi-analytical analysis}\label{sec3}
\begin{figure}[ht]
\begin{center}
\includegraphics[width=8cm]{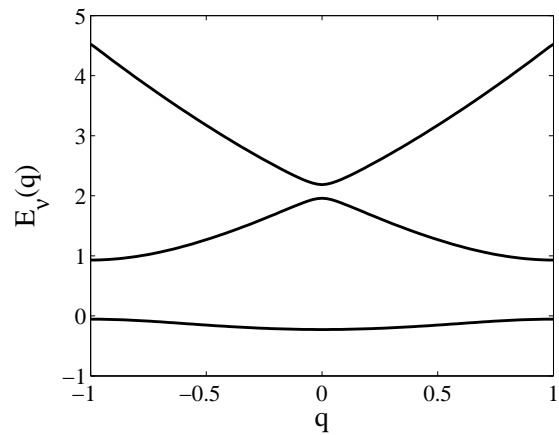}
\caption{The first three Bloch bands $E_\nu^n(q)$ ($\nu=1,\,2,\,3$). The
amplitude $Un=0.5$. All parameters are dimensionless. }
\label{fig1}
\end{center}
\end{figure}

In the previous section we introduced the model Hamiltonian. Furthermore, it was assumed that $L\gg2\pi k^{-1}$, making sure that the system is quasi periodic. Neglecting boundary effects arising from having a finite cavity length, the corresponding eigenvalue problem relaxes to the Mathiue equation
\begin{equation}
\hat{H}_n|\phi_\nu^n(q)\rangle=E_\nu^n(q)|\phi_\nu^n(q)\rangle.
\end{equation}
As a periodic operator, the spectrum
$E_\nu^n(k)$ for given $n$ is characterized by a band index
$\nu=1,2,3,...$ and a quasi momentum $q$ defined within the first Brillouin zone ($-1\leq q<1$). The corresponding eigenstates $|\phi_\nu^n(q)\rangle$ are the so called Bloch functions. A typical spectrum is envisaged in Fig.~\ref{fig1}. The
figure shows the first three energy Bloch bands $E_\nu^n(q)$.

Typically, incoming atoms are not in pure momentum
eigenstates. The atomic velocity selection is never perfect causing an uncertainty in the mean momentum. This is taken into account by considering initial Gaussian states
\begin{equation}\label{inat}
\psi(p,0)=\frac{1}{\sqrt[4]{2\pi\Delta_p^2}}e^{-\frac{(p-p_0)^2}{4\Delta_p^2}},
\end{equation}
where $\Delta_p$ is its width determined by the velocity uncertainty
in the state preparation, and $p_0$ the mean initial momentum. The time evolution of this state is rendered by the Hamiltonian
(\ref{ham2}). For a small coupling $Un$ and moderate
spreading $\Delta_p\ll1$, the atomic state will
predominantly populate a single Bloch band $\nu'$ with average quasi
momentum $q_0=p_0-\nu'$. By expanding the corresponding energy around $q_0$
\begin{equation}
E_{\nu'}^n(q)\approx E(q_0)+v_g^n(q_0)(q-q_0)+\frac{1}{2}\frac{1}{m_n^*(q_0)}(q-q_0)^2,
\end{equation}
where $v_g^n=dE_{\nu'}^n(q)/dq|_{q=q_0}$ and
$m_n^*(q_0)=\left(d^2E_{\nu'}^n(q)/dq^2\right)^{-1}$ it follows that, for a
given $n$, the time evolved state in $x$-representation approximates~\cite{effmass}
\begin{equation}
\psi_n(x,t)\approx\frac{1}{\sqrt[4]{8\pi\Delta_p^2\Delta_x^4}}e^{-\frac{(x-v_g^nt)^2}{4\Delta_x^2}},
\end{equation}
where
\begin{equation}
\Delta_x^2=\frac{1}{4\Delta_p^2}+\frac{it}{2m_n^*},
\end{equation}
and $v_g^n\equiv v_g^n(q_0)$ and $m_n^*\equiv m_n^*(q_0)$.
Within these approximations, the wave packet preserves its Gaussian
form, moves with a group velocity $v_g^n$, and spreads according
to the effective mass $m_n^*$.

\begin{figure}[ht]
\begin{center}
\includegraphics[width=8cm]{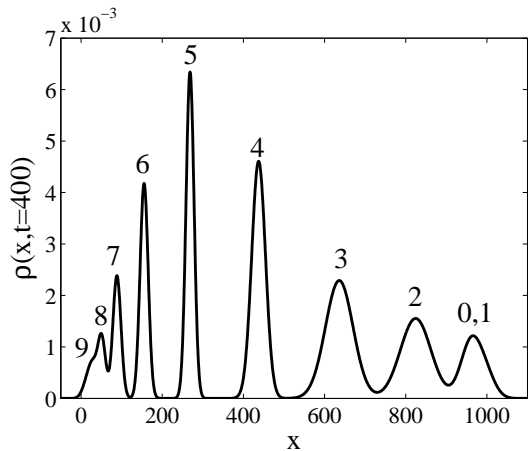}
\caption{The atomic density (\ref{dens}) at time $t=400$. The field
is initially in a coherent state with $\bar{n}=4$ and the inset numbers indicate the corresponding number of photons. The parameters are $p_0=2.58$, $\Delta_p=p_0/50$, and $U=0.7$. The initial velocity $p_0$ implies that the atomic wave packet evolves on the third Bloch band. }
\label{fig2}
\end{center}
\end{figure}

For a general initial state of the cavity field
$|\phi\rangle=\sum_{n=0}^\infty c_n|n\rangle$, we obtain the single-band 
approximated time evolved atom-field state
\begin{equation}\label{timestate1}
\Psi(x,t)\approx\sum_{n=0}^\infty \psi_n(x,t)c_n|n\rangle,
\end{equation}
and the corresponding atomic density
\begin{equation}\label{dens}
\rho_{at}(x,t)=\sum_{n=0}^\infty|c_n|^2|\psi_n(x,t)|^2.
\end{equation}
Due to the uncertainty of photon numbers, it follows that the atomic state will split into a set of Gaussians. Hence, whenever the distances between consecutive Gaussian wave packets exceeds their widths, a measurement of the atomic position will reveal the photon number $n$. Equivalent to a position measurement is a time-of-flight measurement, where the time it takes the atom to traverse the cavity is recorded. The atomic density $\rho_{at}(x,t=400)$ is depicted in
Fig.~\ref{fig2}. The initial cavity field is a coherent state
\begin{equation}
c_n=e^{-|\alpha|^2/2}\frac{\alpha^n}{\sqrt{n!}}
\end{equation}
with an average number of photons $\bar{n}=|\alpha|^2=4$. For the
current set of parameters, the wave packet evolves on the third
Bloch band. The inset numbers give the corresponding number of photons $n$. For $\bar{n}=4$, the population of the Fock state with $n=10$ is less than one percent and therefore only the first ten Fock states are seen. 

\begin{figure}[ht]
\begin{center}
\includegraphics[width=8cm]{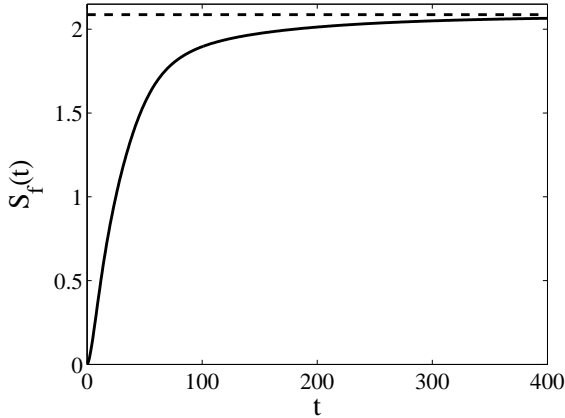}
\caption{Time evolution of the von Neumann entropy (\ref{entropy}). Parameters are the
same as in Fig.~\ref{fig2}. The amount of entanglement is rapidly
increasing and asymptotically reaches its maximum value
$S_f^{(max)}$ indicated by the dashed line. } \label{fig3}
\end{center}
\end{figure}

For the example of Fig.~\ref{fig2}, a single atomic detection is most likely sufficient for determining the photon number. However, for shorter interaction times the Gaussians overlap and a single measuremnt cannot resolve the photon number. Nonetheless, by repeating the measurement procedure for a second and third atom and so on, the photon distribution will finally pick a single Fock state $|n\rangle$ despite the non-zero Gaussian overlaps. Naturally, the less resolved the peaks are the more atoms are needed. Stated in other words, a strong atom-field correlation implies fewer atomic detections. Considering pure initial states and a closed system, we employ the von Neumann entropy as an estimate of the amount of atom-field correlation/entanglement. For the reduced field density operator $\rho_f(t)=\mathrm{Tr}_{at}[\rho(x,t)]$, where $\rho(x,t)$ is the full system density operator and the trace is over the atomic motional degrees of freedom, we have the von Neumann entropy
\begin{equation}\label{entropy}
S_f(t)=-\mathrm{Tr}_f\big[\rho_f(t)\ln[\rho_f(t)]\big].
\end{equation}
The trace is over field degrees of freedom. The
maximum entropy is given by
$S_f^{(max)}=-\sum_n|c_n|^2\ln[|c_n|^2]$, with $c_n$ the initial photon
amplitudes. Using the same parameters as in Fig.~\ref{fig2}, we
present the corresponding entropy in Fig.~\ref{fig3}. The dashed
line gives the maximum entropy, and it is clear that $S_f(t)$
approaches this value in the long time limit. On the other hand, for times $t$ less than say 100, a series of atomic detection is most likely required for an efficient QND measurement.

\section{Numerical analysis}\label{sec4}
\subsection{Complications due to a realistic system}
In realistic experimental setups, several complicating effects arise that were not fully addressed in the previous section. In this section we will take them into account numerically. 

By entering the cavity, we assume that the atom feels a fairly sudden turn on of the dipole induced cavity potential. To model the potential (\ref{pot1}) for the whole $x$-axis, we introduce an envelope function such that
\begin{equation}
V(\hat{x})\!=\!\frac{U}{2}\!\left[\tanh\!\left(\!\frac{\hat{x}\!-\!L/2}{X_s}\right)\!-\tanh\!\left(\!\frac{\hat{x}\!+\!L/2}{X_s}\right)\!\right]\!,
\end{equation}
where $X_s$ determines the slope of the envelope function around $x=\pm L/2$ and we naturally chose $X_s\ll L$. The rapid change in $V(\hat{x})$ in the vicinity of $x=\pm L/2$ will induce some backward scattering of both the incoming and outgoing atomic wave packets. Thereby, especially for large amplitudes $Un$, part of the wave packet will not reach the detector. Experimentally, backward scattered atoms may be ignored and we will focus only on the forward scattered atoms, those who are detected. The measured time $t_n$ is presumably the time between atomic state preparation and the detection. As no backwarded scattered atoms are recorded, $t_n$ includes the free space time propagation $t_{fr}$ plus the time $\tau_n$ spent inside the cavity; $t_n=\tau_n+t_{fr}$. Since the atom-field interaction is dispersive, the atomic velocity before and after the cavity are the same and independent of $n$, and consequently $t_n$ is an indirect measure of the group velocity $v_g^n(q_0)\approx L/\tau_n$.

Yet another complication is the fact that the true state of the atom (\ref{inat}) is not restricted to a single Bloch band $\nu'$. From the symmetry of the Hamiltonian, and given $n$, it follows that the atomic state expressed in Bloch functions is
\begin{equation}\label{blochin}
\psi(p,0)=\sum_\nu\int_{-1}^{+1}d_\nu^n\psi(q,0)|\phi_\nu^n(q)\rangle dq.
\end{equation}
Here, $\psi(q,0)$ is $\psi(p,0)$ with the real momentum replaced by the quasi momentum~\cite{com1}, and the $d_\nu^n$ are the proper weights given $Un$. For small $Un$, the coefficient $|d_{\nu'}^n|\approx1$ with $\nu'=p_0-q_0$. For the parameters employed in the previous section one finds that all $|d_{\nu'}^n|$ is noticeable smaller than 1, and the atomic wave packet will therefore split up into sub-packets as it enters the cavity. The different parts deriving from atomic propagation according to the corresponding Bloch bands. 

It is clear that it is a trade-off between having a large separation of group velocities $v_g^n$, and small atomic scattering and splitting. The first favors large amplitudes $U$, while effects originating from scattering and splitting are decreased for small values of $U$. Despite this, we will now demonstrate that by repeated measurements, the efficiency can be made asymptotically close to unity.         

\subsection{Results}   
The Schr\"odinger equation (\ref{ham1}) is solved using the split-operator method~\cite{split}. The atmic intial condition is a Gaussian with width $\Delta_x=15\gg2\pi k^{-1}$, initial position $x_0=-L/2-70$, and initial momentum $p_0=3.75$. The cavity length $L$ is varied, while $X_s=0.2$ and $U=0.7$ are kept fixed throughout. The initial cavity field is as before, {\it i.e.} a coherent state with $\bar{n}=4$. Note that the atomic wave packet is initialized in the regime where the amplitude of the cavity field is approximately zero; $V(-L/2-70)\approx0$. The time propagation $t_f$ is performed till the various fowardly scattered atomic wave packet components have left the interaction region. At this instant $t_f$, due to the back scattering we renormalize $\rho_{at}(x,t_f)$ accordingly 
\begin{equation}
\rho_{at}'(x,t_f)\equiv\left\{\begin{array}{ll}
0, & x<L/2 \\
\frac{\rho_{at}(x,t_f)}{\int_{L/2}^\infty\rho_{at}(x,t_f)dx}, & x\geq L/2.
\end{array}\right.
\end{equation} 

\begin{figure}[ht]
\begin{center}
\includegraphics[width=8cm]{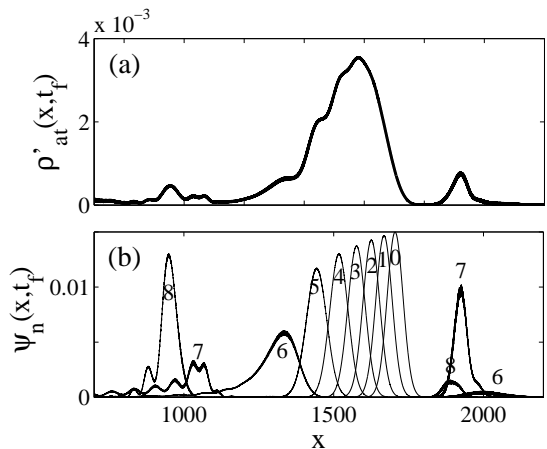}
\caption{The renormalized atomic density $\rho_{at}'(x,t_f)$ (a) and the different atomic wave packet components $\psi_n'(x,t_f)$ (b). The cavity length $L=1400$ and final time $t_f=660$. Other dimensionless parameters are $\Delta_x=15$, $x_0=-770$, $p_0=3.75$, $\bar{n}=4$, $U=0.7$, and $X_s=0.2$.} \label{fig4}
\end{center}
\end{figure}

In Fig.~\ref{fig4} (a) we present the renormalized atomic density for $x\geq L/2$. The cavity length for this example is fairly long, $L=1400$. Compared with the ideal situation of Fig.~\ref{fig2}, we note that in this more realistic situation the various atomic wave packet components $\psi_n(x,t_f)$ are less resolved from each other. In Fig.~\ref{fig4} (b) we give the corresponding components $\psi_n'(x,t_f)$ renormalized within the interval $x\in[L/2,\infty]$. The numbers represent the photon number $n$. The peaks around $x\approx2000$, corresponding to photon numbers $n=6,\,7,\,8$, originate from the wave packet component propagating on higher excited Bloch bands. Expectedly, this component becomes significantly populated only for strong field amplitudes $Un$. 

For the example of Fig.~\ref{fig4}, a single atomic measurement will not reveal the photon number with very high efficiency. However, as mentioned in the previous section, repeated atomic measurements will improve the scheme considerably. First we note that numerically, instead of considering a time measurement we can equally well freeze the time evolution and make a position measurement. After the first atom has been recorded with a corresponding position $x_r$, the photon distribution becomes $|c_n^{(1)}|^2=|\psi_n'(x_r,t_f)|^2|c_n|^2/N$, where $N=\sum_n|c_n^{(1)}|^2$. For the second atom traversing the cavity, its atomic density $\rho_{at}'^{(1)}(x,t)$ is given by Eq.~(\ref{dens}) with $c_n$ replaced by $c_n^{(1)}$. Each atomic measurement acts as a ``photon filter"~\cite{jonas1}, modifying the photon distribution with the weights $\psi_n'(x_r,t_f)$. 

For the numerical simulation, the position $x_r$ of atom $j$ is randomly picked according to the corresponding probability distribution $\rho_{at}'^{(j)}(x,t_f)$. Once the position has been determined, the photon distribution is adjusted accordingly and a new atomic density for atom $j+1$ is calculated. The process is repeated until only a single photon component survives the filtering. The number of iterations needed depends on; the randomly picked numbers and on how well separated the atomic wave packet components are. In general, large $L$ implies less atomic measurements. 
 
\begin{figure}[ht]
\begin{center}
\includegraphics[width=8cm]{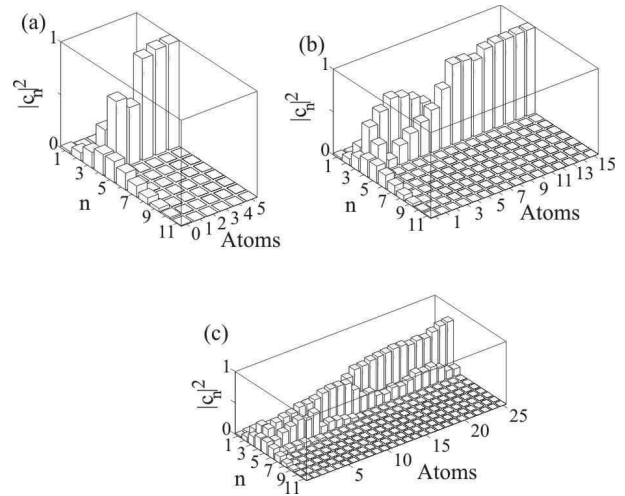}
\caption{Simulations of the QND measurement using successive atomic measurements. The three plots correspond to $L=1400$, $L=600$, and $L=200$ respectively. For longer cavities, less atomic measurements are required to have an efficient QND measurement. The rest of the parameters are the same as in Fig.~\ref{fig4} except $x_0=-L/2-70$. } \label{fig5}
\end{center}
\end{figure}

The results of three simulations for different cavity lengths $L$ are presented in Fig.~\ref{fig5}. In the first case (a), the parameters are the same as for Fig.~\ref{fig4}. Already after three atomic measurements the photon distribution has collapsed to approximately a single Fock state. For the cases with $L=600$ (b) and $L=200$ (c), considerably more atoms are needed in order to single out a photon number.  

\begin{figure}[ht]
\begin{center}
\includegraphics[width=8cm]{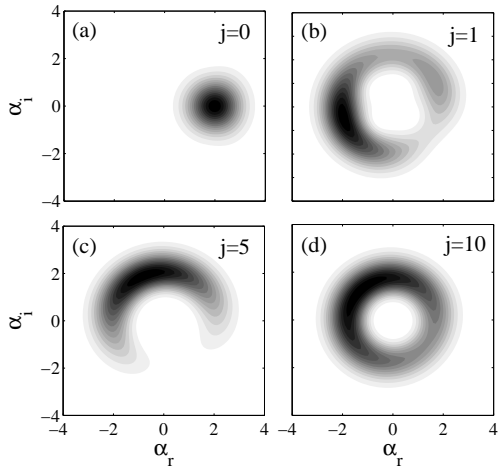}
\caption{Evolution of the $Q$-function (\ref{qfun}) after the positions of $j=0,\,1,\,5,\,10$ atoms have been recorded. For the initial state (a), the $Q$-function is a Gaussian centered around $(\alpha_r,\alpha_t)=(2,0)$. Even after 1 atomic measurement (b), the phase space quasi distribution does show a circular structure characterizing the Fock number state. After 5 measurements (c), the circular structure has actually declined, while after 10 atoms (d) the distribution is almost perfectly circular with a radius $|\alpha|\approx1.7$ corresponding to the $n=3$ Fock state. The parameters are as in Fig.~\ref{fig5} (b), but the sequence of random numbers $x_r$ is not the same as those used for that plot.} \label{fig6}
\end{center}
\end{figure}

Each atomic measurement is projective, leaving the field in a pure state $|\phi\rangle_j=\sum_n\psi_n'^{(j)}(x_r,t_f)c_n^{(j)}|n\rangle/N_j$,  $N_j$ being the proper normalization constant. The filtering projection onto a single Fock state seen in Fig.~\ref{fig5} can also be demonstrated via the phase space distributions, {\it e.g.} the Husimi $Q$-function~\cite{mandel}
\begin{equation}\label{qfun}
Q^{(j)}(\alpha)=\frac{\langle\alpha|\phi\rangle_{\!j\,j}\!\langle\phi|\alpha\rangle}{\pi}.
\end{equation}
Here, $|\alpha\rangle$ is a coherent state with amplitude $\alpha$. For a coherent state, the $Q$-function is Gaussian, while for a Fock state it is a circle with radius $|\alpha|=\sqrt{n}$. Figure~\ref{fig6} gives four examples of $Q^{(j)}(\alpha)$ for $j=0$ (a), $j=1$ (b), $j=5$ (c), and $j=10$ (d). The initial state and the parameters are the same as those of Fig.~\ref{fig5} (b). In plots (b) and (c), there are two photon numbers dominating, $n=4,\,5$. In (d), almost all population resides in the $n=3$ state.

\section{Discussion and concluding remarks}\label{sec5}
\begin{figure}[ht]
\begin{center}
\includegraphics[width=8cm]{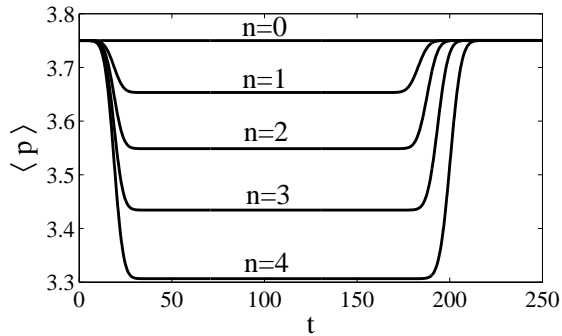}
\caption{The average momentum for an atomic wave packet traversing the $L=600$ cavity. The different $n$'s give the number of photons. The parameters are the same as those of Fig.~\ref{fig5} (b). } \label{fig7}
\end{center}
\end{figure}

The idea behind our scheme is different from that of Ref.~\cite{qndquant1}. In~\cite{qndquant1}, due to virtual exchange of photons with the cavity field, atoms are deflected perpendicularly with respect to their initial velocity. A position measurement is therefore an indirect measure of the number of photons that has been exchanged. Even for short interaction times (imposing the Raman-Nath approximation), the atom can acquire a certain number of $2\hbar k$ momentum kicks, where the factor 2 comes from the fact that the interaction is dispersive and absorption of a photon by the atom is always accompanied by emittance of a photon. The measurement then projects onto any of the momentum eigenstates $p=2r\hbar k$, $r=0,\,1,\,2,\,...\,$. In the present scheme, on the other hand, the time-of-flight measurement indirectly gives the wave packet velocity which for weak couplings approximate the group velocity $v_g$, characterizing the average velocity $\langle\hat{p}\rangle/m$ for the corresponding Bloch state. To make these arguments more transparent, in Fig.~\ref{fig7} we display the time evolution of $\langle\hat{p}\rangle$ for the $L=600$ cavity and $n=0,\,1,\,2,\,3,\,4$. The plot makes clear that the velocity decreases for increasing photon numbers, and also that the atom regains its initial velocity after exiting the interaction region. 

The measurement efficiency is enhanced by considering long cavities. However, long cavities naturally implies long interaction times and cavity losses may become significant. For short cavities, the individual atomic interaction times are shorter but, on the other hand, more atomic measurements are needed. This again might cause long total operational times. One way of decreasing the total process time is by using feedback-techniques~\cite{feedback}. Nonetheless, even when cavity losses become important the scheme can be useful. For a lossy cavity, one typically considers an external pumping of the cavity, keeping the field in a coherent state with amplitudes determined from balancing the pump and loss rates. If the time-scale for the field is much shorter than that of repeated atomic measurements, the field attains its steady state between each measurement. The result of the measurements will then reveal the steady state photon distribution. 

As a summary, we have introduced a QND measurement scheme of the photon numbers in a cavity. It relies on time-of-flight measurements of atoms traversing a Fabry-Perot cavity along its axial direction. The field intensity $n$ directly affects the atomic velocity while traversing the cavity; the velocity drop is increased for larger photon numbers $n$. This can be explained using the language of group velocities for particles moving in periodic potentials. Employing this concept, we argued that for a sufficiently long cavity a single atom can resolve the photon number non-destructively. By decreasing the cavity length, repeated atomic measurements are required. Our analytical results were numerically verified considering more realistic cavity QED setups. In particular, we found that typically more than single atomic measurements are needed for reliable QND measurements. We further explained that even for lossy cavities the scheme is of interest as it can be used as a tool to measure the full steady state photon distribution. Getting access to the full state of the field would require more sophisticated approaches like tomography measurements~\cite{tomo}.

\begin{acknowledgments}
JL acknowledges support from the MEC program (FIS2005-04627).
\end{acknowledgments}


\begin{thebibliography}{999}
\bibitem{com1} For the scaled parameters, the quasi momentum $q$ is related to the real momentum $p$ via $q=\mod(p,1)$ and the band index is $\nu=p-q$.
\end{thebibliography}
\end{document}